\documentclass[usegraphicx,useAMS,usenatbib]{mn2e}

\pubyear{2004}

\begin{document}

\title[Towards a Precision Cosmology from Starburst Galaxies at $z > 2$]
{Towards a Precision Cosmology from Starburst Galaxies at $z > 2$}

\author[E. R. Siegel et al.]
{E. R. Siegel,$^{1}$\thanks{E-mail: siegel@phys.ufl.edu} R. Guzm\'an,$^{2}$ Jorge P. Gallego,$^{2}$ M. Ordu\~na L\'opez$^{2}$ and
\newauthor P. Rodr\'\i guez Hidalgo$^{2}$ \\
$^1$Department of Physics, University of Florida, 
Gainesville, FL, 32611-8440, USA \\
$^2$Department of Astronomy, University of Florida, Gainesville, FL, 32611, USA}

\date{Accepted ???? ??????? ??.  Received ???? ??????? ??; in original form 2004 June 18}

\maketitle

\begin{abstract}
This work investigates the use of a well-known empirical correlation between the velocity dispersion, metallicity, and luminosity in H$\beta$ of nearby H$\rm{II}$ galaxies to measure the distances to H$\rm{II}$-like starburst galaxies at high redshifts.  
This correlation is applied to a sample of 15 starburst galaxies with redshifts between $z=2.17$ and $z=3.39$ to constrain $\Omega_m$, using data available from the literature.  A best-fit value of $\Omega_m=0.21^{+0.30}_{-0.12}$ in a $\Lambda$-dominated universe and of $\Omega_m=0.11^{+0.37}_{-0.19}$ in an open universe is obtained.  A detailed analysis of systematic errors, their causes, and their effects on the values derived for the distance moduli and $\Omega_m$ is carried out.  A discussion of how future work will improve constraints on $\Omega_m$ by reducing the errors is also presented.
\end{abstract}

\begin{keywords}
cosmological parameters -- 
HII regions --
galaxies: starburst -- 
galaxies: high redshift -- 
galaxies: fundamental parameters --
distance scale
\end{keywords}

\section{Introduction}

Precision cosmology, or accurately constraining the parameters describing the universe, has recently become an active field of research due to the precision of available data sets.  Stringent contraints have recently been placed on cosmological parameters from measurements of the microwave background \citep{B03}, type Ia supernovae \citep{R00}, and galaxy surveys \citep{H03,Ba03}.  Although these sources of data are sufficient for generating consistent values for the mass density ($\Omega_m$), vacuum energy density ($\Omega_\Lambda$), the dark energy equation of state parameter ($w$), and the value of spatial curvature in the universe ($\Omega_k$), these values must be checked via as many independent methods as possible for consistency, accuracy, and avoiding systematic biases.  Furthermore, without the data from supernovae, there would be weak evidence at best for stating that $w \approx -1$, thus it is important to seek another, independent observation supporting the existence of dark energy.

The cosmological parameter with the greatest number of observable cross-checks is $\Omega_m$.  It has been derived using many techniques, including the Sunyaev-Zel'dovich effect \citep{G01}, weak gravitational lensing \citep{H01}, X-ray luminosities \citep{B01}, large scale clustering \citep{SBCG03}, peculiar velocities of galaxy pairs \citep{F03}, and supernovae data \citep{K03}.  These methods yield results ranging from $\Omega_m = 0.13$ to $\Omega_m = 0.35$, and are all consistent with one another at the 2-$\sigma$ level.  However, they all face difficulties when attempting to differentiate between cosmological models, as they are only weakly dependent on $\Omega_\Lambda$, $\Omega_k$, and $w$.  If a reliable standard candle were found at high redshifts, cosmological models could be discriminated between by precise and accurate observations, as the distance modulus becomes sensitive to $\Omega_\Lambda$, $\Omega_k$, and $w$ at higher redshifts.  It is known that local H$\rm{II}$ galaxies and giant H$\rm{II}$ regions in local galaxies are physically similar systems \citep{M87}.  This work extrapolates a link between nearby H$\rm{II}$ galaxies and H$\rm{II}$-like starburst galaxies at high redshifts to use such objects as standard candles.  This is accomplished through the application of the known correlation between the luminosity in the $ H\beta $ line ($L_{H\beta}$), the velocity dispersion ($\sigma$), and metallicity ($O/H$) of nearby H$\rm{II}$ galaxies \citep[discovered by][hereafter MTM]{MTM88} to the H$\rm{II}$-like starburst galaxies found at high redshifts.  This correlation, when applied to starburst galaxies at $z > 2$, allows for discrimination between different values of $\Omega_m$ \citep[as first suggested by][hereafter MTT]{MTT00}, and can discern which cosmological model is most favored by the data.

H$\rm{II}$ galaxies (and H$\rm{II}$ regions of galaxies) are characterized by a large star-forming region surrounded by singly ionized hydrogen.  The presence of O- and B-type stars in an H$\rm{II}$ region causes strong Balmer emission lines in $H\alpha$ and $H\beta$.  The size of a giant H$\rm{II}$ region was shown to be correlated with its emission line widths by \citet{M78}.  This correlation was improved upon by \citet{TM81}, who showed that $L_{H\beta}$ of giant H$\rm{II}$ regions is strongly correlated with their $\sigma$.  This basic correlation, its extension to H$\rm{II}$ galaxies, and its usefulness as a distance indicator have been explored in the past \citep[MTM;][]{MTM87,M87}.  The empirical correlation for H$\rm{II}$ galaxies (MTM) relates their $L_{H\beta}$, $\sigma$, and $O/H$.  The relationship is
\begin{equation}
\log{L_{H\beta}}=\log{M_z} + 29.60 \mathrm{,} \qquad M_z \equiv \frac{\sigma^5}{O/H} \mathrm{,}
\end{equation}
where the constant 29.60 is determined by a zero-point calibration of nearby giant H$\rm{II}$ regions (MTT) and from a choice of the Hubble parameter, $H_0 = 71 \, km \, s^{-1} \, Mpc^{-1}$ \citep{H02,F01}.  The 1-$\sigma$ rms scatter about this correlation is $\pm 0.33 \, \, \rm{dex}$ on $\log{L_{H\beta}}$ from the local sample of H$\rm{II}$ galaxies found in MTM.  Starburst galaxies observed at high redshifts exhibit the same strong Balmer emission lines and intense star formation properties \citep{P01,E03} as nearby H$\rm{II}$ galaxies.  This work follows the suggestion of MTT that equation 1 applies to the H$\rm{II}$-like starburst galaxies found at high redshifts, and provides evidence to validate that assumption.

The remainder of this paper discusses the constraints that can be placed on $\Omega_m$ and the restrictions that can be placed on the choice of cosmology using starburst galaxies.  Section 2 details how the data set was selected and analyzed.  Section 3 states the results obtained from the selected data.  Section 4 provides a discussion of the random and systematic errors associated with this method, including a detailed discourse on the assumption of universality between local H$\rm{II}$ galaxies and high redshift starburst galaxies.  Finally, section 5 presents the conclusions drawn from this work, and points towards useful directions for future work on this topic.

\section{Selection of the Data Sample}

The goal of the analysis presented here is to obtain distances for each H$\rm{II}$-like starburst galaxy at high redshift.  H$\rm{II}$ galaxies must first be detected at high redshift.  A sample is then selected on the basis of the correlation in equation 1 holding and for which the distance modulus ($DM$) can be computed from the observed quantities.  The quantities required for analysis of these galaxies are $\sigma$, the flux in $H\beta$ ($F_{H\beta}$), $O/H$, the extinction in $H\beta$ ($A_{H\beta}$), and the equivalent width in the $H\beta$ line ($EW$).


Following the analysis in MTT, the distance modulus of H$\rm{II}$ galaxies can be derived from:
\begin{equation}
DM=2.5\log(\frac{\sigma^5}{F_{H\beta}})-2.5\log({O/H})-A_{H\beta}-26.18 \mathrm{,}
\end{equation}
where the constant $26.18$ is determined by $H_0$ and equation 1.  This paper makes extensive use of equation 2 because it expresses $DM$ purely in terms of observables.  $DM$ is insensitive to $\Omega_m$, $\Omega_\Lambda$, $\Omega_k$, and $w$ at low redshifts ($z \leq 0.1$), differing by $0.1$ magnitudes or fewer for drastic changes in the choice of parameters above.  At high redshifts ($z > 2$), however, $DM$ can vary by up to $3$ magnitudes depending on the choice of parameters.  Of the four parameters above available for variation, $DM$ is most sensitive to changes in $\Omega_m$, as noted by MTT.  However, for values of $\Omega_m \leq 0.3$, $DM$ is sensitive to variations in the other parameters by $0.2$ to $0.5$ magnitudes.  Since other measurements indicate that indeed $\Omega_m \leq 0.3$, this paper also considers variations in $\Omega_\Lambda$ and $\Omega_k$.


Data for starburst galaxies at $z > 2$ are found in \citet{P01} and \citet{E03}, which contain measurements for many of the desired observables (and related quantities), along with redshift data.  Partial measurements exist for 36 starburst galaxies.  
According to MTT, the correlation in equation 1 holds true for young H$\rm{II}$ galaxies whose dynamics are dominated by O- and B-type stars and the ionized hydrogen surrounding them.  As H$\rm{II}$ galaxies evolve in time, short-lived O- and B-stars burn out quickly.  Although some new O- and B-stars are formed, eventually the death rate of O- and B-stars exceeds their birth rate, causing a galaxy to be under-luminous in $H\alpha$ and $H\beta$ for its mass.  This effect can be subtracted out by examining the $EW$ of these galaxies, and cutting out the older, more evolved galaxies (those with smaller equivalent widths).  For this paper, a cutoff of $EW > 20 \, {\rm\AA}$ is adopted, and galaxies with $EW \leq 20 \, {\rm\AA}$ are not included, similar to the cutoff advocated in MTT.  
There are also galaxies with large $EW$ that do not follow the correlation of equation 1 within a reasonable scatter.  
It is well-known that a large fraction of local H$\rm{II}$ galaxies contain multiple bursts of star formation \citep{MTM87}.  If multiple unresolved star-forming regions are present, the observed $\sigma$ will be very large due to the relative motion of the various regions.  Such galaxies are not expected to follow the correlation of equation 1, as articulated in \citet{MTM87}.  The simplest way to remove this effect is to test for non-gaussianity in the emission lines from this effect, but signal-to-noise and resolution are insufficient to observe this effect.  Since $\sigma$ for a system of multiple star-forming regions will be much higher than for a single H$\rm{II}$ galaxy, a cut can be placed on $\sigma$ to remove this effect.  Monte Carlo simulations indicate that if $\sigma$ is observed to be greater than $130 \, km \,s^{-1}$, it is likely due to the presence of multiple star-forming regions.  To account for this presence, all galaxies with $\sigma > 130 \, km \,s^{-1}$ are discarded.
Imposing the cuts on $\sigma$ and $EW$ selects 15 of the 36 original galaxies, creating the data sample used for the analysis presented here.

\begin{table*}
\begin{minipage}{160mm}
\caption{Selected starburst galaxies with their properties and $DM$.}
\begin{center}
\begin{tabular}{@{}cccccccc}
Galaxy & $z$ & $\sigma$ \footnote{$\sigma$ is given in units of $km \, s^{-1}$}
 & $F_{H\beta}$ \footnote{$F_{H\beta}$ is given in units of $10^{-17} \, erg \, s^{-1} \, cm^{-2}$}
 & 12+$\log{(O/H)}$ & $ A_{H\beta} $ \footnote{$A_{H\beta}$ is given in mag} 
 & $EW$ \footnote{$EW$ is given in units of ${\rm\AA}$}
 & $DM$ \footnote{$DM$ is given in mag with 1-$\sigma$ random errors.}\\ 
Q0201-B13 & 2.17 & $62 \pm 29$ & $0.9 \pm 0.2$ & 8.55 & 0.013 & 23 & $47.49^{+2.10}_{-3.43}$ \\
Q1623-BX432 & 2.18 & $51 \pm 22$ & $1.9 \pm 0.5$ & 8.55 & 0.157 & 72 & $45.45^{+1.97}_{-3.07}$ \\
Q1623-MD107 & 2.54 & $ \leq 42 $ & $1.3 \pm 0.3$ & 8.55 & 0.141 & 21 & $44.82^{+0.31}_{-1.58}$ \\
Q1700-BX717 & 2.44 & $ \leq 60 $ & $1.3 \pm 0.3$ & 8.55 & 0.285 & 25 & $46.64^{+0.31}_{-1.58}$ \\
Q1700-MD103 & 2.32 & $75 \pm 21$ & $2.4 \pm 0.6$ & 8.55 & 0.735 & 47 & $46.72^{+1.38}_{-1.80}$ \\
SSA22a-MD41 & 2.17 & $107 \pm 15$ & $2.6 \pm 0.7$ & 8.55 & 0.214 & 31 & $48.96^{+0.78}_{-0.85}$ \\
CDFa C1 & 3.11 & $ \leq 63 $ & $3.4 \pm 1.0$ & 8.55 & 0.505 & 28 & $45.77^{+0.31}_{-1.58}$ \\
Q0347-383 C5 & 3.23 & $69 \pm 4$ & $\leq 1.7$ & 8.55 & 0.237 & $\leq 27$ & $47.12^{+0.44}_{-0.32}$ \\
B2 0902+343 C12 & 3.39 & $87 \pm 12$ & $2.7 \pm 0.3$ & $8.70 \pm 0.08$ & 0.773 & 37 & $46.96^{+0.71}_{-0.81}$ \\
Q1422+231 D81 & 3.10 & $116 \pm 8$ & $4.1 \pm 0.4$ & $8.62 \pm 0.07$ & 0.237 & 43 & $48.81^{+0.38}_{-0.40}$ \\
SSA22a-MD46 & 3.09 & $67 \pm 6$ & $\leq 2.3$ & 8.55 & 0.110 & $\leq 31$ & $46.76^{+0.56}_{-0.51}$ \\
SSA22a-D3 & 3.07 & $113 \pm 7$ & $1.3 \pm 0.3$ & $8.39 \pm 0.16$ & 1.01 & 25 & $49.71^{+0.43}_{-0.41}$ \\
DSF2237+116a C2 & 3.32 & $100 \pm 4$ & $3.5 \pm 0.4$ & 8.55 & 0.852 & 25 & $47.73^{+0.25}_{-0.25}$ \\
B2 0902+343 C6 & 3.09 & $55 \pm 15$ & $3.0 \pm 1.0$ & 8.55 & 0.284 & 40 & $45.22^{+1.38}_{-1.76}$ \\
MS1512-CB58 & 2.73 & 81 & $1.35 \pm 0.2$ & $8.49 \pm 0.10$ & 1.14 & 26 & $47.49^{+1.22}_{-1.57}$ \\
\end{tabular}
\end{center}
\end{minipage}
\end{table*}

Once the sample has been selected, the quantities required to calculate $DM$ using equation 2 must be tabulated for each selected galaxy.  Not all of the necessary data are available in the literature for these galaxies, so assumptions have been made to account for the missing information.  $z$ was measured for all galaxies by the vacuum heliocentric redshifts of the nebular emission lines.  $\sigma$ was obtained for all galaxies from the broadening of the Balmer emission lines, $H\alpha$ for the galaxies from \citet{E03} and $H\beta$ for the galaxies from \citet{P01}.  $F_{H\beta}$ is measured directly for the galaxies in \citet{P01}, but \citet{E03} measures $F_{H\alpha}$ instead, thus $F_{H\alpha}$ must be converted to $F_{H\beta}$.  The conversion for emitted flux is given by \citet{O89} as $F_{H\alpha} = 2.75 \, F_{H\beta}$, but observed fluxes must correct for extinction.  Thus, the complete conversion from $F_{H\alpha}$ to $F_{H\beta}$ will be given by
\begin{equation}
F_{H\beta} = \frac{1}{2.75} \, F_{H\alpha} \, 10^{(\frac{A_{H\alpha}-A_{H\beta}}{2.5})}  \mathrm{,}
\end{equation}
where $A_{H\alpha}$ and $A_{H\beta}$ are the extinctions in $H\alpha$ and $H\beta$, respectively.  Obtaining $O/H$ is more difficult, as measurements of metallicity only exist for 5 of the 36 original starburst galaxies.  An average value of $O/H$ is used for the galaxies where $O/H$ measurements are unavailable.  
\begin{figure}
 \includegraphics[width=8cm]{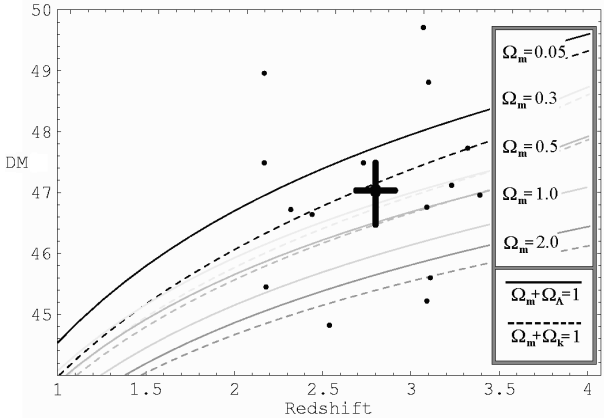}
\caption{$DM$ vs. $z$ for various cosmological models, with the selected galaxies.  The crosshairs represents the 1-$\sigma$ constraints on the $DM$ vs. $z$ parameter space from the selected data sample.}
\label{fig1}
\end{figure}
Values of $O/H$ are obtained through measurement of the $[$O$ \rm{\, II}]$ emission line at $ 3727 \, {\rm\AA}$ and the $[$O$ \rm{\, III}]$ lines at $ 4959 \, {\rm\AA}$ and $5007 \, {\rm\AA}$ for five of the galaxies in \citet{P01}.  The strong line index $R_{23}$ \citep{P79} is assumed to have its temperature-metallicity degeneracy broken towards the higher value of $O/H$, as is shown to be the case for luminous starburst galaxies at intermediate redshifts in \citet{KK00}.  The combination of the oxygen line measurements with this assumption yields values for $O/H$ for these galaxies.  The mean value of $O/H$ is then taken to be the average metallicity for each of the other galaxies where such line measurements are unavailable.  \citet{S04}, using the $[$N$ \rm{\, II}]$/$H\alpha$ ratio as their metallicity indicator, obtains an average $O/H$ of 8.33 for the galaxies found in \citet{E03}.  This value is noted as a possible improvement to the one chosen here, and is further discussed in section 4 of this work.
$A_{H\beta}$ is derived from the $E(B-V)$ color of the galaxy in question.  Extinction laws are known and established for the Milky Way, the Large and Small Magellanic Clouds (LMC and SMC, respectively), and the H$\rm{II}$ regions of the LMC and SMC \citep{G03}, but have not been established for starburst galaxies in general.  This paper assumes dust in H$\rm{II}$ galaxies to be comparable to that in giant H$\rm{II}$ regions, thus $A_{H\beta}$ for starburst galaxies is taken to be the $A_{H\beta}$ derived in \citet{G03} for the H$\rm{II}$ regions of the LMC and SMC.  A best fit applied to the data in \citet{G03} yields $A_{H\beta} = (3.28 \pm 0.24) \, E(B-V)$ and $A_{H\alpha} = (2.14 \pm 0.17) \, E(B-V)$ for starburst galaxies.  These results are also applicable to the flux conversion in equation 3.  $E(B-V)$ is unavailable for the galaxies from \citet{P01}, but can be derived by noticing the correlation between $E(B-V)$ and corrected ($G-R$) colors for starburst galaxies in \citet{E03}.  The conversion adopted is $E(B-V) \approx 0.481 \, (G-R)$.  Finally, $EW$ is measured for all galaxies in \citet{P01}, but \citet{E03} gives only the spectra for the $H\alpha$ line.  $EW$ is estimated for the \citet{E03} galaxies by estimating the continuum height from each spectra and the area under each $H\alpha$ peak, calculating the equivalent width in $H\alpha$, and converting to $H\beta$ using the Balmer decrements of \citet{O89}.  The complete data set is listed in Table 1.

\section{Results}

In the previous section, the distance modulus was calculated for each galaxy in the selected sample.  The comparison of these values of $DM$ and the predicted values of $DM$ at a given redshift for different cosmological models provides a constraint on the cosmological parameters.  $DM$ is most sensitive at high redshifts to the variation of the cosmological parameter $\Omega_m$, as pointed out by MTT.  $\Omega_m$ is therefore the parameter which is constrained most tightly by observations of starburst galaxies.  Each galaxy yields a measurement for $DM$ and for $z$.  Although there are multiple models consistent with each individual measurement, observations of many galaxies at different redshifts will allow the construction of a best-fit curve, which is unique to the choice of cosmological parameters $\Omega_m$, $\Omega_\Lambda$, $\Omega_k$, and $w$.  The data sample of 15 galaxies in this paper is insufficient to distinguish between models in this fashion, as the uncertainties in each individual measurement of $DM$ are too large.  The method by which the uncertainties can be reduced is to bin the data according to redshift and find a best-fit value of $DM$ at that point.  Due to the size of the sample in this paper, all 15 points are 
averaged into one point of maximum likelihood to constrain the cosmology, with errors arising from the random errors of the individual points and from the distribution of points.  The average value obtained is $DM = 47.03^{+0.46}_{-0.56}$ at a redshift $ z = 2.80 \pm 0.11 $.  The different cosmological models, along with the most likely point and the raw data points, are displayed in figure 1, with $H_0 = 71 \, km \, s^{-1} \, Mpc^{-1}$.

The constraints placed on $\Omega_m$ from this analysis are $ \Omega_m=0.21^{+0.30}_{-0.12} $ in a $\Lambda$-dominated universe ($\Omega_m + \Omega_\Lambda = 1$; $\Omega_k=0$) and $ \Omega_m=0.11^{+0.37}_{-0.19} $ in an open universe ($\Omega_m + \Omega_k = 1$; $\Omega_\Lambda=0$).  Figure 2 shows the comparison in $\Omega_m$ vs $\Omega_\Lambda$ parameter space between the preliminary constraints of this work and early constraints arising from CMB data and SNIa data \citep[from][]{dB00}.  

\begin{figure}
 \includegraphics[width=8cm]{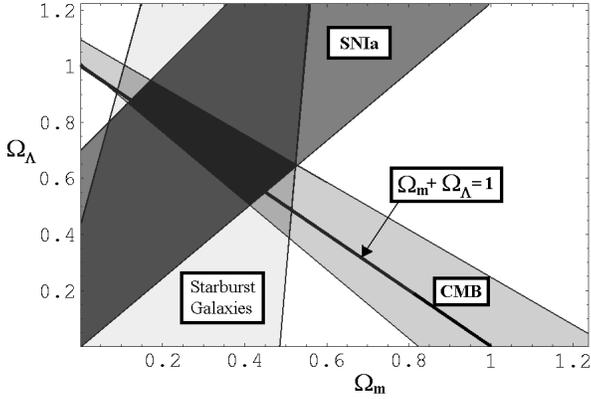}
\caption{1-$\sigma$ constraints in $\Omega_m$ vs. $\Omega_\Lambda$ parameter space from starburst galaxies, along with older constraints from CMB and SNIa data, found in \citet{dB00}.}
\label{fig2}
\end{figure}

CMB and SNIa constraints led to the first reliable estimates of $\Omega_m$ and $\Omega_\Lambda$.  The preliminary constraints presented here are comparable to early constraints from CMB and SNIa data, as shown in figure 2.  The accuracy in $\Omega_m$ and $\Omega_\Lambda$, as determined from the most recent CMB and SNIa data \citep{B03} is now $\pm 0.04$ in each parameter.  A similar, and perhaps even superior accuracy can be achieved using starburst galaxies at high redshifts, as discussed in the next two sections.

\section{Discussion}

There have been many assumptions made along the path to obtaining $\Omega_m$ via the method described above.  Each assumption carries along with it an associated error.  Some of the assumptions made are inherent to the method used, while others affect only the data sample specific to this paper.  Both will lead to systematic errors, although the sample-specific errors will largely be eliminated by improved measurements.  Additionally, random errors result from uncertainties in the measurements and from the intrinsic scatter in the distribution of points.  An analysis of all three types of errors ensues below.

The most important assumption made was the assumption of universality of the distance indicator used for both local H$\rm{II}$ galaxies and H$\rm{II}$-like starburst galaxies.  Support for this assumption is provided by the fact that both galaxy types follow the empirical correlation of equation 1, as shown in figure 3.  
The validity of the correlation between $L_{H\beta}$ and $M_z$ can be tested directly to determine its range of applicability.  
By assuming a cosmology, $\log{L_{H\beta}}$ can be written purely in terms of luminosity distance ($d_L$), $F_{H\beta}$, and $A_{H\beta}$, which are either measurable or computable from observables for each galaxy.  $\log{M_z}$ can be determined through measured values for $\sigma$ and $O/H$.  Comparing the quantities $\log{L_{H\beta}}$ and $\log{M_z}$ then allows a test of the correlation in equation 1 for all galaxies of interest.  All available H$\rm{II}$ and H$\rm{II}$-like starburst galaxies with appropriately measured quantities are included to test the correlation.  Local galaxies are taken from MTM and from the Universidad Complutense de Madrid (UCM) survey \citep{G96,V96,Zpc}, intermediate redshift starburst galaxies are taken from \citet{G97}, and high redshift starburst galaxies are from \citet{P01} and \citet{E03}.  The cosmology assumed to test universality is $\Omega_m=0.3$, $\Omega_\Lambda=0.7$, and cuts are applied to all samples so that $EW > 20 \, {\rm\AA}$ and $\sigma < 130 \, km \, s^{-1}$.  The results are shown in figure 3.




The major reasons to conclude that the assumption of universality is valid lie in figure 3.  There exists an overlap between all four samples in both $L_{H\beta}$ and $M_z$, from the sample where the correlation is well established (nearby samples, MTM and UCM), to intermediate redshift H$\rm{II}$-like starburst galaxies \citep{G97}, to the high redshift sample used in this paper, from \citet{E03} and \citet{P01}.  These four samples all follow the same correlation between $L_{H\beta}$ and $M_z$ within the same intrinsic scatter (although the observed scatter broadens at high redshifts due to measurement uncertainties).  The samples are all consistent with the same choice of slope and the same choice of zero-point.  For these reasons, the correlation of equation 1 appears 
to be just as valid for H$\rm{II}$-like starburst galaxies as for nearby H$\rm{II}$ galaxies.  

It is important to note that there is an uncertainty in the zero-point calibration of figure 3 of $\pm 0.08 \, \, \rm{dex}$, which has not improved since MTM (1988).  This corresponds to an uncertainty in $DM$ of $\pm 0.20$.  If starburst galaxies are to be taken seriously as a distance indicator for precision cosmology, it is essential that the zero-point be determined to significantly greater accuracy.  This can be accomplished via a comprehensive survey of the nearby ($z < 0.1$) H$\rm{II}$ galaxy population, significantly increasing the sample size from the MTM sample.

\begin{figure}
 \includegraphics[width=8cm]{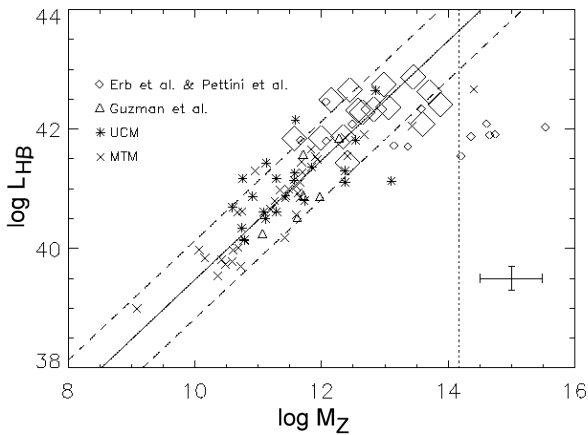}
\caption{$\log{M_z}$ vs. $\log{L_{H\beta}}$ for local H$\rm{II}$ galaxies and starburst galaxies at intermediate and high redshifts.  The solid line is the best fit of the correlation to the local data set, flanked by the dashed lines, which give the 2-$\sigma$ rms scatter.  The large diamonds represent the selected high redshift data sample; the small diamonds are the data not selected on the basis of either $EW$ or $\sigma$.  The vertical dotted line is the derived cut on $\sigma$ of $130 \, km \, s^{-1}$.  The crosshairs represents the typical uncertainty in each selected data point.}
\label{fig3}
\end{figure}

The other assumptions which are inherent to this method are the choices of where to cut on $EW$ and on $\sigma$, and the assumption that $A_{H\beta}$ is the same for starburst galaxies as it is for local H$\rm{II}$ regions.  Moving the $EW$ cut from $EW > 20 \, {\rm\AA}$ up to $EW \geq 25 \, {\rm\AA}$, as suggested in MTT, would systematically raise the $DM$ by $0.14$ mag for this sample.  The $EW$ threshold for the onset of major evolutionary effects is not yet well-determined.  
The cut on $\sigma$ comes about in order to remove contamination from objects containing multiple unresolved star-forming regions.  Since the correlation between $L_{H\beta}$ and $M_z$ is only valid for single H$\rm{II}$ galaxies and H$\rm{II}$ regions \citep{MTM87}, a cut must be made to remove objects containing multiple star-forming regions.  Single H$\rm{II}$ galaxies have a gaussian distribution in $\sigma$ peaked at $\sim 70 \, km \, s^{-1}$, but objects with multiple unresolved regions have an entirely different distribution.  A cut on $\sigma$ at $130 \, km \, s^{-1}$ retains $95$ per cent of the valid, single H$\rm{II}$ galaxies, while eliminating $75$ per cent of the contaminating objects.  Additionally, it can be shown that the contaminating objects which are not eliminated depart only slightly from the empirical correlation of equation 1.  
It is essential, for any future survey, that the proper cuts be determined and applied to $EW$ and $\sigma$, otherwise substantial uncertainties will arise from the selection of the data sample.  Finally, the derived $A_{H\beta}$ itself has an uncertainty of $\pm 38$ per cent, due to the fact that there are competing extinction laws that give different results \citep{G03,C94,C00}.  Both laws are comparably grey, but have different normalizations.  The difference between the two laws leads to a systematic uncertainty in the $DM$ of $\pm 0.17$ mag.  

There have also been assumptions made specifically to compensate for incomplete data in the \citet{P01} and \citet{E03} data sets.  The systematic uncertainties that these assumptions induce can be eliminated in future surveys through measurements of all required quantities.  The assumption that the temperature-metallicity degeneracy is most likely broken towards the upper branch of the $R_{23}$ vs. $O/H$ curve for luminous starburst galaxies at high redshift is based on sound analysis \citep{KK00}, but is still a dangerous one to make.  Measurement of the $4363 \, {\rm\AA}$ oxygen line would break the $R_{23}$ vs. $O/H$ degeneracy, and yield a definite value for metallicity for each galaxy.  Furthermore, $O/H$ had to be assumed for 11 of the 15 galaxies in the sample, inducing a possible systematic which could affect $DM$ if the assumed average $O/H$ differs from the true value.  If the value for $O/H$ from \citet{S04} is used for the galaxies selected from \citet{E03}, the average $DM$ is raised by 0.22 mag.  This systematic can be removed in future surveys by a measurement of the $[$O$\, \rm{II}]$ line at $3727 \, {\rm\AA}$ and the $[$O$\, \rm{III}]$ lines at $4959 \, {\rm\AA}$ and $5007 \, {\rm\AA}$ for each galaxy.  Measurement of the $[$N$\, \rm{II}]$ line at $6584 \, {\rm\AA}$, along with $H\alpha$, can provide another measurement of metallicity \citep{S04}.  Many abundance indicators are available \citep{KD02}, and future surveys should allow multiple, independent techniques to be used, significantly reducing errors.  
There is a large uncertainty on the order of $\pm 30$ per cent in the measurement of $EW$ for the \citet{E03} sample due to the difficulty of establishing the height of the continuum.  Some galaxies may have been included which should not have been, and others may have been excluded which should have been present.  The effect on the distance modulus is estimated to be $\pm 0.16$ mag, but this will be removed by measuring equivalent width in $H\beta$ with a higher signal-to-noise spectra for all galaxies in future surveys.  Finally, $E(B-V)$ colors, a substitute for $A_{H\beta}$ measurements, are unavailable for galaxies from \citet{P01}, and were derived from an approximate correlation noticed between $E(B-V)$ and the corrected $(G-R)$ colors in \citet{E03}.  There is an overall uncertainty in the extinction due to the fact that the average derived extinction for the \citet{E03} and the \citet{P01} samples differ by $0.34 \, \, \rm{dex}$.  Thus, there is an induced systematic in $DM$ of $0.17$ mag, which will be eliminated when $A_{H\beta}$ measurements are explicity taken for all galaxies.

Random errors, due to both uncertainties in measurement and to the large scatter in the distribution of points, are perhaps the best understood of the sources of error.  Measurements of $A_{H\beta}$ are uncertain by $0.04$ to $0.11 \, \, \rm{dex}$, depending on the galaxy's brightness.  Improved measurements, which rely on the $H\alpha / H\beta$ ratio instead of solely on $E(B-V)$ colors, may reduce the uncertainty significantly.  Measurements of $F_{H\beta}$ are uncertain by roughly $20$ to $25$ per cent on average, and random uncertainties in $O/H$ are of order $0.10 \, \, \rm{dex}$.  The largest measurement uncertainty comes from measurements of $\sigma$, which is obtained by the broadening of the Balmer emission lines.  Even relatively small uncertainties in $\sigma$ of order $15$ per cent can induce uncertainties in $DM$ of $0.8$ mag per galaxy.  The induced uncertainty is so large because, as seen in equation 2, $DM$ is dependent on $\sigma^5$, whereas it depends only linearly on the other quantities.  Future work will be able to measure the $H\alpha$ and $H\beta$ lines, as well as three oxygen lines, which should improve the measurements of $\sigma$, further reducing the random uncertainties.  The distribution of points may not improve as statistics improve due to the intrinsic scatter on the $M_z$ vs. $L_{H\beta}$ relation, but random errors all fall off as the sample size increases.  The errors decrease as $N^{-1/2}$, where $N$ is the number of galaxies in the sample.  Even if random errors associated with intrinsic properties (such as $F_{H\beta}$, $\sigma$, or $O/H$) remain large for individual galaxies, increasing the sample size will drive down the overall random errors.  Hence, a sample of 500 galaxies, as opposed to 15, will have its random uncertainties reduced by a factor of 6 or better.  The new generation of Near-IR Multi-Object Spectrographs (such as FLAMINGOS and EMIR) in 10 meter class telescopes will be ideal for obtaining all necessary measurements for such a sample. 

\section{Conclusions}


This paper has demonstrated that using H$\rm{II}$-like starburst galaxies at high redshift as a standard candle is a promising and well-motivated avenue to explore for precision cosmology.  A future survey of high redshift starburst galaxies with measurements of $z$, $\sigma$, $A_{H\beta}$, $F_{H\beta}$, $O/H$, and $EW$ will reduce both random and systematic errors dramatically.  Since the inherent scatter of the method is large, a large sample size is required to obtain meaningful constraints.  This paper contains a sample size of only 15 galaxies, but future surveys should be able to obtain hundreds of starburst galaxies that survive the selection cuts.  For a sample of 500 galaxies, this will improve constraints on $\Omega_m$ to a restriction of $\pm 0.03$ due to random errors.  Additionally, all of the systematics specific to this sample due to incomplete data will disappear.

If the assumption of universality between local H$\rm{II}$ galaxies and high redshift starburst galaxies is correct, this method of measuring $\Omega_m$ is capable of providing very tight constraints, independent of any constraints arising from other sources, including CMB and SNIa data.  
Additionally, if galaxies are obtained at a variety of redshifts between $2 \leq z \leq 4$, different cosmological models (including vacuum-energy dominated models with different values of $w$) can be tested for consistency with the future data set.  If $\Omega_m \leq 0.3$, the differences in $DM$ at various redshifts become quite pronounced, and meaningful results as to the composition of the non-matter components of the universe can be obtained as well.  
Future work on this topic has the potential to provide strong independent evidence either supporting or contradicting the concordance cosmological model of $\Omega_m + \Omega_\Lambda = 1$, $w=-1$, in addition to providing a very stringent constraint on the $\Omega_m + \Omega_\Lambda$ parameter space.

\section*{Acknowledgments}

The authors thank the University of Florida Alumni Fellowship Program for support.  R.G. acknowledges funding from NASA grant LTSA NAG5-11635.  Extensive use has been made of NASA's ADS bibliographic services, and of the software packages Mathematica and IDL.  We thank Brian Cherinka for assistance with computing extinction laws, Jesus Gallego and Jaime Zamorano for providing unpublished data from their survey of local H$\rm{II}$ galaxies, and Jim Fry, Chip Kobulnicky, and David Koo, for their critical insights during the early stages of this work.  Finally, we thank Mariano Moles for his informative and constructive comments on our work, and the anonymous referee, for a thoroughly helpful report on the first version of this paper.


\end{document}